# INFERRING THE PARTIAL CORRELATION STRUCTURE OF ALLELIC EFFECTS AND INCORPORATING IT IN GENOME-WIDE PREDICTION

Carlos Alberto Martínez[1], Kshitij Khare[2], Syed Rahman[2], Mauricio A. Elzo[1]
[1]Department of Animal Sciences
[2]Department of Statistics
University of Florida, Gainesville, FL, USA

## SUMMARY

In this study, we addressed the problem of genome-wide prediction accounting for partial correlation of marker effects when the partial correlation structure, or equivalently, the pattern of zeros of the precision matrix is unknown. This problem requires estimating the partial correlation structure of marker effects, that is, learning the pattern of zeros of the corresponding precision matrix, estimating its non-null entries, and incorporating the inferred concentration matrix in the prediction of marker allelic effects. To this end, we developed a set of statistical methods based on Gaussian concentration graph models (GCGM) and Gaussian directed acyclic graph models (GDAGM) that adapt the existing theory to perform covariance model selection (GCGM) or DAG selection (GDAGM) to genome-wide prediction. Bayesian and frequentist approaches were formulated. Our frequentist formulations combined some existing methods with the EM algorithm and were termed Glasso-EM, CONCORD-EM and CSCS-EM, whereas our Bayesian formulations corresponded to hierarchical models termed Bayes G-Sel and Bayes DAG-Sel. Results from a simulation study showed that our methods can accurately recover the partial correlation structure and estimate the precision matrix. Methods CONCORD-EM and Bayes G-Sel had an outstanding performance in estimating the partial correlation structure and a method based on CONCORD-EM yielded the most accurate estimates of the precision matrix. Our methods can be used as predictive machines and as tools to learn about the covariation of effects of pairs of loci on a given phenotype conditioned on the effects of all the other loci considered in the model. Therefore, they are useful tools to learn about the underlying biology of a given trait because they help to understand relationships between different regions of the genome in terms of the partial correlations of their effects on that trait.

**Key words**: Concentration graph models, Correlated marker allelic effects, Directed acyclic graph models, Inverse covariance model selection, Partial correlation networks.

## 1 INTRODUCTION

Since early stages of genome-wide prediction (Meuwissen et al., 2001), it has been known that marker allelic effects on a given phenotype may be correlated (Gianola et al., 2003). So far, approximations to account for correlated marker effects have involved imposing an arbitrary covariance structure based on the idea that only nearby markers are correlated (Gianola et al., 2003; Yang and Tempelman, 2012). Accounting for correlation among marker effects in genome-wide prediction involves the estimation of covariance matrices in high dimensional settings which is a problem of contemporary interest in statistics (Letac and Massam, 2007; Khare and Rajaratnam, 2011). It is frequently assumed that the matrix to be estimated is sparse, that is, many of its entries are zero. Recently, the application of graphical models in this complex problem has shown to be a useful and flexible approach to find sparse estimators of covariance or precision (also known as concentration matrix) matrices (Carvalho et al., 2007; Letac and Massan, 2007; Rajaratnam et al., 2008) and this area has experienced a great expansion. An undirected graph $G$ encodes the pattern of zeros either in the covariance (covariance graph models) or the precision matrix (concentration graph models) of a set of random variables corresponding to the nodes of $G$, hence the term "graphical models". When the joint distribution of these random variables is assumed to be multivariate Gaussian and sparsity is imposed in the precision matrix, the model is known as a Gaussian concentration graph model (GCGM). In this case the model is said to be Markov with respect to the undirected graph $G$, that is, variables not sharing an edge in $G$ are conditionally independent given all the remaining variables (Dawid and Lauritzen, 1993; Letac and Massam, 2007; Ben-David et al., 2015).

In submitted research (Martínez et al., 2016a; Martínez et al., 2016b) we proposed methods based on graphical models to account for marginally or partially correlated marker effects. There, it was assumed that the graph dictating the pattern of zeros in the covariance or precision matrix was known because it could be built using domain-specific knowledge. However, it is of interest to develop methods for inverse covariance estimation when the pattern of zeros is unknown. In the context of GCGM this sort of problem is known as covariance selection, sparse covariance selection, sparse graphical model selection or partial correlation selection and it implies estimating the pattern of zeros and the non-null entries of the precision matrix (Friedman et al., 2008; Bickel and Levina 2008; Peng et al., 2009; Rajaratnam et al., 2008; Khare et al., 2015). This allows carrying out genome-wide prediction accounting for partially correlated marker effects and finding the partial correlation structure of these effects for a particular phenotype. Alternatively, sparsity can be induced in the Cholesky parameter of the concentration matrix which is equivalent to selecting an underlying directed acyclic graph (DAG) model (Shojaie and Michailidis, 2010; Ben-David et al., 2015; Khare et al., 2016).

Frequentist and Bayesian methods have been proposed to perform covariance estimation when the pattern of zeros is unknown. On the frequentist side, there are several approaches based on penalized likelihood or pseudo-likelihood functions. On the other hand, one of the main features of the Bayesian approach is the computational burden; therefore, some hybrid strategies combining frequentist and Bayesian approaches have been proposed (Ben-David et al., 2015).

However, all these methods have been developed in the context of datasets with identically and independently distributed (IID) observations from a distribution whose covariance/concentration matrix needs to be sparsely estimated. The objective of this study was to develop statistical methods adapting the theory of GCGM and Gaussian DAG models to the more complex setting of genome-wide prediction in order to find the partial correlation structure of the allelic effects of molecular markers on a phenotype and to take it into account in the prediction of additive genetic values.

## 2 MATERIALS AND METHODS

This section is organized as follows. Because GCGM and Gaussian DAG models theory is not widespread in the realm of quantitative and statistical genetics, in the first part of this section some methods to perform graphical model selection/DAG selection are briefly described. Then, the challenge encountered when using this theory in genome-wide prediction and the approaches to overcome it are presented. Finally, a simulation study designed to test the ability of the proposed methods to recover the partial correlation structure and to estimate the concentration matrix is described.

### 2.1 Inverse Covariance Model Selection/DAG Selection
#### 2.1.1 Some frequentist approaches

A popular method to perform variable selection at the level of the regression coefficients in a linear model is the Lasso (Tibshirani, 1996). Using the regression interpretation of the elements of the precision matrix, the ideas in Lasso can also be applied to sparse inverse covariance estimation. Let $Y_1, Y_2, \ldots, Y_N$ be independent random vectors identically distributed $MVN(\mathbf{0}, \Omega^{-1})$ and let $\boldsymbol{\theta}^j = \left(\frac{-\Omega_{ji}}{\Omega_{jj}}\right)_{i \neq j} \in \mathbb{R}^{p-1} \; \forall \; j = 1, 2, \ldots, p$, then $\boldsymbol{\theta}^j = \underset{\beta \in \mathbb{R}^{p-1}}{\mathrm{argmin}} E[X_j - \boldsymbol{\beta}' \boldsymbol{Y}_{-j}]$ where $\boldsymbol{Y}_{-j} = (Y_k)_{k \neq j}$. Using this property, Meinshausen and Buhlmann (2006) proposed a method to find the sparsity pattern based on the Lasso. They proposed the following estimator: $\widehat{\boldsymbol{\theta}}_j = \underset{\beta \in \mathbb{R}^{p-1}}{\mathrm{argmin}} \left\{ \frac{1}{n} \|\boldsymbol{Y}^{(j)} - \sum_{k \neq j} \beta_k \boldsymbol{Y}^{(k)}\|^2 + \lambda \|\boldsymbol{\beta}\|_1 \right\}$ where $\boldsymbol{Y}^{(j)} = (Y_j^i)_{i=1}^n$, $1 \leq j, k \leq p$. This method has the limitation that the estimated concentration matrix is not symmetric; therefore, although it was a seminal study, this method is not frequently used nowadays. Friedman et al. (2008) proposed the graphical Lasso (GLasso) approach, which also takes the Lasso idea from regression problems to perform sparse inverse covariance estimation. This method poses an $l_1$ penalty over the off-diagonal elements of the concentration matrix. The function to be minimized is as follows: $Q_{GL}(\Omega) = tr(\Omega S) - log|\Omega| + \lambda \sum_{1 \leq i < j \leq p} |\Omega_{ij}|$. A feature of this function is that if the sample covariance matrix $S$ is singular, then the function is not strictly convex and, as a result, a unique global minimum may not exist. An alternative to the existing methods was proposed by Peng et al. (2009) and it is known as the SPACE (Sparse PArtial Correlation Estimation) algorithm. This method uses a pseudo-likelihood approach, takes the symmetry of the problem into account, and involves the minimization of an objective function involving an $l_1$ penalty over the partial correlation coefficients between pairs of variables given all the remaining variables. It has the advantage that it does not explicitly assume normality. However, it has the drawback that its objective function is not jointly convex in the elements of $\Omega$ but biconvex, that is, when a part of the function is fixed (a set of variables) the function is convex in the remaining part, and therefore, its convergence properties are not well established. A solution to these limitations was proposed by Khare et al. (2015) and it is known as the convex correlation selection method and algorithm (CONCORD). They developed a convex formulation of this pseudo-likelihood graphical model selection problem, the following is the objective function of CONCORD: $Q_{CONCORD}(\rho, \Omega) = \frac{n}{2}\sum_{k=1}^{p}\mathbf{\Omega}'_{.k}S\mathbf{\Omega}_{.k} + \lambda \sum_{1 \leq k < j \leq p}|\rho^{kj}| - n\sum_{k=1}^{p}\log\Omega_{kk}$, where $\mathbf{\Omega}_{.k}$ is the $k^{th}$ column of $\Omega$ and $\rho^{kj} = \frac{-\Omega_{kj}}{\sqrt{\Omega_{kk}\Omega_{jj}}}$ is the partial correlation coefficient between variables $k$ and $j$ given all other variables. Because $\sum_{k=1}^{p}\mathbf{\Omega}'_{.k}S\mathbf{\Omega}_{.k}$ is a positive semidefinite quadratic from in $\Omega$ and $-\log x$ and $|x|$ are both convex functions, $Q_{CONCORD}(\Omega)$ is a jointly convex function. A coordinate-wise minimization algorithm was proposed by Khare et al. (2015) in order to minimize $Q_{CONCORD}(\rho, \Omega)$; the computational complexity was found to be $\min\{O(np^2), O(p^3)\}$.

The convex formulation of the problem permitted to prove the convergence of the CONCORD algorithm even if $N < p$ (Khare et al., 2015). Because the interest was in estimating the sparsity pattern of the inverse covariance matrix, Khare et al. (2015) did not restrict the solution to be positive definite. Hence, the parameter space is $M := \{\Omega \in \mathcal{S}, \Omega_{kk} > 0 \ \forall \ k = 1,2, \ldots, p\}$, where $\mathcal{S}$ is the set of real symmetric matrices. In this algorithm, the function $Q_{CONCORD}$ is minimized with respect to every $\Omega_{kj}$ for $1 \leq k, j \leq p$; holding all other entries fixed. The problem has the following closed form solutions. For $1 \leq k \leq p$, $Q_{CONCORD}$ is minimized at $\Omega_{kk} = \frac{-(\sum_{l \neq k}\Omega_{kl}S_{kl}) + \sqrt{(\sum_{l \neq k}\Omega_{kl}S_{kl})^2 + 4S_{kk}}}{2S_{kk}}$, and for $1 \leq k < j \leq p$, $Q_{CONCORD}$ is minimized at $\Omega_{kj} = \frac{S_{\frac{\lambda}{n}}\left(-\left(\sum_{j' \neq j}\Omega_{kj'}S_{jj'} + \sum_{k' \neq k}\Omega_{k'j}S_{kk'}\right)\right)}{S_{kk}+S_{jj}}$ where $S_\eta(x) = sgn(x)(|x| - \eta)_+$ is the soft thresholding operator, $(z)_+$ is the positive part of $z$. Therefore, a single iteration of the coordinate-wise minimization algorithm for CONCORD involves updating each entry of $\Omega$ based on the above expressions.

Another approach to estimate the concentration matrix when the pattern of zeros is unknown is to induce sparsity in its Cholesky parameter. Huang et al. (2006) proposed a non-parametric method known as the Cholesky Lasso, which is based on the interpretation of elements of the Cholesky parameter of the inverse covariance matrix as the coefficients of a sequence of linear regressions that allows implementing the Lasso. Following the same idea, when variables have a natural ordering, Levina et al. (2008) derived a method known as nested LASSO. Recently, Khare et al. (2016) developed the convex sparse Cholesky selection (CSCS) method which induces sparsity through an $l_1$ penalty on the conventional Cholesky parameter of $\Omega$. By doing so, the objective function is jointly convex and the estimator is guaranteed to be positive definite. In addition, this method has convergence guarantees even when $N < p$ (Khare et al., 2016). This method also requires an ordering of the variables.

### 2.1.2 Bayesian approaches

A conjugate prior that corresponds to the Diaconis-Ylvisaker prior (Diaconis and Ylvisaker, 1979) is the G-Wishart (GW) distribution (Roverato, 2000). It is similar to the inverse Wishart distribution except that its support is the following cone: $\mathbb{P}_G = \{A: A \in \mathbb{P}^+ \text{ and } A_{ij} = 0 \text{ whenever } (i,j) \notin E\}$, where $\mathbb{P}^+$ is the space of positive definite matrices. The un-normalized density of the GW distribution is: $\pi_{(U,\delta)}(\Omega) \propto |\Omega|^{\delta/2} \exp(-tr(\Omega U)/2), U \in \mathbb{P}^+, \delta > 0, \Omega \in \mathbb{P}_G$. A natural method to perform Bayesian model selection is posing a prior over $G$ and then computing: $\pi(G|Y_1, Y_2, \ldots, Y_N) \propto f(Y_1, Y_2, \ldots, Y_N|G)\pi(G)$, i.e., the posterior probability of graph (model) $G$. Under a uniform prior it amounts to compute:

$$f(Y_1, Y_2, \ldots, Y_N|G) = \int_{\mathbb{P}_G} f(Y_1, Y_2, \ldots, Y_N|\Omega)\pi(\Omega|G)d\Omega \qquad (1)$$

The model with the largest value for this marginal likelihood is selected. Although this is a conceptually simple approach, computational burden is an issue. For a $p$-dimensional problem, that is, $|V| = p$, there are $2^{p(p-1)/2}$ possible undirected graphs which grows exponentially with $p$. In addition, in some cases the marginal likelihood presented in Equation 1 cannot be found in closed form; therefore, numerical methods have to be used. This integral can be computed in closed form for decomposable graphs (Letac and Massam, 2007, Rajaratnam et al., 2008).

On the other hand, Ben-David et al. (2015) developed a family of priors called the DAG-Wishart. Let $\mathcal{D}$ be a DAG, then a Gaussian DAG model over $\mathcal{D}$ ($\mathcal{N}_\mathcal{D}$) is defined as the collection of all multivariate Gaussian distributions obeying the directed Markov property with respect to $\mathcal{D}$, that is, if $(y_1, \ldots, y_p) \sim MVN(\mathbf{0}, \Omega^{-1})$ and parents are given smaller indices than their children, then $MVN(\mathbf{0}, \Omega^{-1}) \in \mathcal{N}_\mathcal{D}$ if $y_i \perp \mathbf{y}_{\{1,\ldots,i-1\}\setminus pa(i)} | \mathbf{y}_{pa(i)}$ where $pa(i)$ is the set of parents of node $i$ (Ben-David et al., 2015), $\forall i = 2, \ldots, p$. For a Gaussian model in $\mathcal{N}_\mathcal{D}$, the structure of the DAG $\mathcal{D}$ is reflected in the modified Cholesky decomposition of $\Omega$ which has form $LD^{-1}L$ where $L$ is a lower triangular matrix with diagonal elements equal to 1 and $D$ is a strictly positive diagonal matrix. Formally, let $L_\mathcal{D} = \{L: L \text{ is lower triangular}, L_{ii} = 1 \ \forall \ 1 \leq i \leq p \text{ and } L_{ij} = 0, i > j, j \notin pa(i)\}$, then, $MVN(\mathbf{0}, \Omega^{-1}) \in \mathcal{N}(\mathcal{D})$ if and only if $L \in L_\mathcal{D}$. Thus, choosing a sparsity pattern in the Cholesky parameter of $\Omega$ is equivalent to choosing an underlying DAG model. The DAG-Wishart family has un-normalized pdf $\pi_{U,\alpha}(L, D) \propto \exp\left(-\frac{1}{2}tr(LD^{-1}L')U\right) \prod_{i=1}^{p} D_{ii}^{-\alpha_i/2}, L \in L_\mathcal{D}, D \in \mathbb{D}$ where $\mathbb{D}$ is the space of diagonal matrices with positive diagonal elements. If $\alpha_j - |pa(j)| > 2$, then its normalizing constant is:

$$Z_\mathcal{D} = \prod_{j=1}^{p} \frac{\Gamma\left(\frac{\alpha_j - |pa(j)|}{2} - 1\right) 2^{\frac{\alpha_j}{2}-1} \pi^{\frac{|pa(j)|}{2}} |U^{<j}|^{\frac{\alpha_j - |pa(j)| - 3}{2}}}{|U^{\leq j}|^{\frac{\alpha_j - |pa(j)|}{2} - 1}} \qquad (2)$$

where $\quad U^{<j} = (U_{ik})_{i,k\in pa_j}, U_{j.}^{<} = (U_{jk})_{k\in pa_j}, U_{j.}^{\leq} = (U_{jj}, U_{j.}^{<})$ and $U^{\leq j} = \begin{bmatrix} U^{<j} & U_{.j}^{\leq} \\ (U_{.j}^{\leq})' & U_{jj} \end{bmatrix}$.

Henceforth, the DAG-Wishart family will be denoted as $\mathcal{D}\text{-W}(U, \boldsymbol{\alpha})$. This family is conjugate for $\mathcal{N}(\mathcal{D})$ because if the prior is $\mathcal{D}\text{-W}(\boldsymbol{\alpha}, U)$ and $Y_1, Y_2, \ldots, Y_N$ is an IID sample from a $MVN(\mathbf{0}, \Omega^{-1})$ in $\mathcal{N}(\mathcal{D})$, then the posterior is a $\mathcal{D}\text{-W}(\widetilde{\boldsymbol{\alpha}}, U + NS)$, where $\widetilde{\boldsymbol{\alpha}} = \{\widetilde{\alpha}_i\}_{p\times 1} = \alpha_i + N$. Hence, under a uniform prior for $\mathcal{D}$, finding the posterior probability of a given DAG amounts to compute the normalizing constant of a $\mathcal{D}\text{-W}(\widetilde{\boldsymbol{\alpha}}, U + NS)$ distribution. In order to make the problem computationally tractable, Ben-David et al. (2015) proposed a "hybrid" approach that involves two main steps. In the first one, the penalized likelihood method to perform DAG selection of Shojaie and Michailidis (2010) is used to find a subset of graphs using the available data by varying the penalty parameter. Then, a uniform prior putting positive mass only on this subset is used to perform Bayesian DAG selection as explained above.

## 2.2 Approaches for Inverse Covariance Model Selection/DAG Selection in Genome-Wide Prediction

The following is the linear regression model considered in this study:

$$\boldsymbol{y} = W\boldsymbol{g} + \boldsymbol{e} \qquad (3)$$

where $\boldsymbol{y} \in \mathbb{R}^n$ is an observable vector of response variables (e.g., corrected phenotypes or de-regressed BV), $\boldsymbol{g} \in \mathbb{R}^m$ is an unknown vector of marker allele substitution effects, $\boldsymbol{e} \in \mathbb{R}^n$ is a vector of residuals, $W_{n\times m}$ is an observable matrix whose entries correspond to one to one mappings from the set of possible genotypes for every individual at each marker locus to a subset of the integers, $W = \{w_{ij}\} = \begin{cases} 1, & \text{if genotype} = BB \\ 0, & \text{if genotype} = BA \\ -1, & \text{if genotype} = AA \end{cases}$, $w_{ij}$ is the mapping corresponding to the genotype of the $i^{th}$ individual for the $j^{th}$ marker. The distributional assumptions are: $\boldsymbol{g}|\Omega \sim MVN(\mathbf{0}, \Omega^{-1})$ and $\boldsymbol{e}|\sigma^2 \sim MVN(0, \sigma^2 I)$ which implies that $\boldsymbol{y}|\boldsymbol{g}, W, \sigma^2 \sim MVN(W\boldsymbol{g}, \sigma^2 I)$. Notice that this is not the same problem addressed in standard partial correlation graph/DAG selection. Here, there is a single $n$-dimensional response vector and the target is not estimating the precision matrix of this random vector, but the precision matrix of the unobservable $m$-dimensional vector $\boldsymbol{g}$. In the following sections, methods to cope with this problem are proposed.

### 2.2.1 Frequentist formulation

In order to develop a frequentist approach, we propose a procedure coupling the EM algorithm (Dempster et al., 1977) with the existing theory to perform inverse covariance model selection/DAG selection. The EM algorithm considers the hypothetical situation of being able to observe $\boldsymbol{g}$ (the augmented data) which would allow carrying out maximum likelihood estimation of the parameter of interest using the joint distribution of $\boldsymbol{y}$ and $\boldsymbol{g}$ (the complete likelihood). By properties of maximum likelihood estimators, this estimator is a function of a sufficient statistic of the parameter of interest, which depends on $\boldsymbol{g}$, and because $\boldsymbol{g}$ is not observed, this statistic is replaced by its expected value taken with respect to the conditional distribution of $g$ given $y$. Notice that in the context of the standard covariance estimation problem, having a single $g$ would be equivalent to having a single observation, i.e., $N=1$. Thus, to overcome this issue, it is assumed that there exists heterogeneity of marker effects across families (e.g., full-sibs or half-sibs) as in Gianola et al. (2003). Thus, data is split into $f$ families such that each one has a different vector of marker effects, then $y_i = W_i g_i + e_i, \forall\ i = 1,2,\ldots,f$. It is also assumed that: $g_1,\ldots,g_f$ are IID $MVN(0, \Omega^{-1})$, $e_1,\ldots,e_f$ are independent $MVN(0, \sigma^2 I_{n_i})$ vectors and $Cov(g_i, e_{i'}) = 0, \forall\ 1 \leq i, i' \leq f$, where $n_i$ is the number of observations in family $i$. Let $S_g = \frac{1}{f}\sum_{i=1}^{f} g_i g_i'$, $g^* = (g_1' \cdots g_f')'$, $W^* = Block\ Diag.\{W_i\}_{i=1}^{f}$, and $e^* = y - W^* g^*$, then the negative complete log-likelihood corresponding to the linear model just described is:

$$-l(\sigma^2, \Omega) = constants + \frac{n}{2}\log\sigma^2 - \frac{1}{2}\left(\log|\Omega| - tr(\Omega S_g)\right) + \frac{\|e^*\|_2^2}{2\sigma^2} \qquad (4)$$

The sufficient statistic for $\boldsymbol{\theta} := (\Omega, \sigma^2)$ is $(S_g, e^{*\prime} e^*)$. From properties of multivariate normal distribution it is known that given $y$, $g_1, \ldots, g_f$ are independent with the following distributions: $g_i | y_i \sim MVN\left(K_i^{-1}\frac{W_i' y_i}{\sigma^2}, K_i^{-1}\right)$, where $K_i := \frac{W_i' W_i}{\sigma^2} + \Omega$. Similarly, it follows that $e^* | y \sim MVN(\sigma^2 V^{-1} y, \sigma^2(I - \sigma^2 V^{-1}))$, where $V = W^{*\prime} I_f \otimes \Omega^{-1} W^* + R$. Consequently, $E[S_g | y] = \frac{1}{f}\sum_{i=1}^{f} K_i^{-1}\left[I_m + \frac{1}{(\sigma^2)^2} W_i' y_i y_i' W_i K_i^{-1}\right]$ and $E[e^{*\prime} e^* | y] = \sigma^2 (n - \sigma^2 tr(V^{-1}) + \sigma^2 y' V^{-1} V^{-1} y)$. Thus, the expectation step of the algorithm entails computing $E[S_g | y]$ and $E[e^{*\prime} e^* | y]$, whereas the maximization step involves GCGM theory. Notice that minimizing the function presented in Equation 4 (i.e., maximizing the complete likelihood) does not induce sparsity; therefore, a penalized version of this function is minimized instead. Thus, at iteration $t$, the maximization step involves the optimization of an objective function corresponding to Equation 4 plus the $l_1$ penalty term $\lambda \sum_{1 \leq i < j \leq p} |\Omega_{ij}|$. Notice that this modified objective function is the result of replacing $-\log \pi(g|\Omega)$ by the objective function of G-Lasso; alternatively, the objective function of the CONCORD method can be used to build a different version of the proposed algorithm. Henceforth, these methods will be referred to as GLasso-EM and CONCORD-EM respectively. Regarding the residual variance, in both methods, the maximization step at iteration $t$ involves computing

$(\hat{\sigma}^2)^{(t+1)} = \frac{\hat{q}^{(t)}}{n}, \hat{q}^{(t)} := E[e^{*\prime} e^* | y]\Big|_{\boldsymbol{\theta} = \boldsymbol{\theta}^{(t)}}$. Although the objective functions of these methods do not correspond to the complete likelihood, notice that both use the sufficient statistic $(S_g, e^{*\prime} e^*)$. In GLasso-EM, the objective function is simply the complete likelihood plus a penalty term. As to the CONCORD-EM algorithm, Lemma 5 in Khare et al. (2015) establishes the connection between the standard Gaussian log-likelihood and the objective function of CONCORD. Basically, the CONCORD objective function can be seen as a reparameterization of the Gaussian likelihood, for further details see Lemma 5 and Remarks 1 and 5 of Khare et al. (2015). On the other hand, recall that CONCORD does not guarantee positive definiteness; thus, at the end of each iteration, the diagonal matrix $0.01 I_m$ is added to the estimate of $\Omega$ in order to

ensure positive definiteness. This step is necessary because if the estimate is not of full rank, then $E[S_g|\mathbf{y}]\big|_{\boldsymbol{\theta}=\boldsymbol{\theta}^{(t+1)}}$ cannot be computed unless $\forall\, i=1,2,\ldots,f, W_i'W_i$ is of full rank which is not typically the case. Once the CONCORD-EM algorithm yields an estimator of the sparsity pattern of $\Omega$, some of the existing methods to estimate the concentration matrix given an undirected graph can be used to construct an EM algorithm similar to those presented in this section. In this study, the method proposed by Hastie et al. (2009, pp 631) was used.

**2.2.2 Bayesian formulation**

Hierarchical models are well-suited to this problem; consequently, a hierarchical Bayes approach is taken. The hierarchical formulation of the model considered here is completed by adding the following layers: $\sigma^2 \sim IG\left(\frac{\tau^2}{2},\frac{v}{2}\right)$, $\Omega|G \sim GW(\delta, U)$, $G \sim \pi(G)$, where $IG(\cdot,\cdot)$ denotes the inverse Gamma distribution. As mentioned before, if there are $m$ markers, there are $2^{m(m-1)/2}$ possible undirected graphs. Therefore, in order to reduce the dimension of the space of graphs, that is, the size of the support set of $\pi(G)$, the idea proposed by Ben-David et al. (2015) of combining frequentist and Bayesian approaches is implemented; thus, the stochastic short-gun search (SSS) algorithm used by these authors is considered here. In a first step, CONCORD is run using 15 values of the penalty parameter yielding a set of 15 graphs. Then, a discrete uniform prior is posed on this reduced space of graphs and as explained above, under this uniform prior, finding the posterior probability of a given graph amounts to finding the marginal likelihood

$$f(\mathbf{y}|G) = \int_{\mathbb{P}_G} \pi(\Omega|G)\left(\int_{\mathbb{R}^m}\int_{\mathbb{R}_+} f(\mathbf{y}|\mathbf{g},\sigma^2,W)\pi(\mathbf{g}|\Omega)\pi(\sigma^2)\,d\sigma^2 d\mathbf{g}\right)d\Omega. \qquad (5)$$

For the hierarchical model considered in this study this integral does not have a closed form solution; therefore, it has to be solved numerically or combining numerical methods with analytical approaches like the Laplace approximation. In particular, the Laplace method permits finding an approximate analytical expression of the inner integral $\int_{\mathbb{R}^{mS}}\int_{\mathbb{R}_+} f(\mathbf{y}|\mathbf{g},\sigma^2,W)\pi(\mathbf{g}|\Omega)\pi(\sigma^2)\,d\sigma^2 d\mathbf{g}$. Once this approximation is found and plugged in Equation 5, Monte Carlo Integration (MCI) is used. When $n > m$, using the Laplace approximation coupled with vanilla MCI results in a much faster method to compute $f(\mathbf{y}|G)$ as compared with a full MCI approach (see Discussion section). Having more phenotypic observations than marker loci is not the most common case in genome-wide prediction, but it can be found in certain populations like the US Holstein for the Illumina's 50K chip (CDCB, 2016) or in cases where the interest is in estimating the partial correlation network of a subset of markers that have been preselected on the basis of previous analyses, e.g., a GWAS or a gene set analysis. After deriving the Laplace approximation to the inner integral and plugging it in Equation 5 the problem reduces to find the normalizing constant of a $GW(\delta+1, \widehat{\mathbf{g}}\widehat{\mathbf{g}}' + U)$ distribution, where $\widehat{\mathbf{g}} := (W'W)^{-1}W'\mathbf{y}$. Closed forms of this normalizing constant are available for decomposable graphs (Letac and Massam, 2007, Rajaratnam et al., 2008), but in this case, general graphs have to be considered. For general graphs, different methods have been proposed to find numerical approximations to the normalizing constant of a GW distribution like the MCI approach developed by Atay-Kayis and Massam (2005). On the other hand, for the case $m > n$, this strategy can be implemented after slightly modifying the prior for $\boldsymbol{g}$, but it was not faster than the one using MCI; running times were practically the same (results not shown). For details on these derivations see the Appendix. Hereinafter this "hybrid" method will be referred to as Bayes G-Sel. Once graph $G$ has been selected, a point estimate of $\Omega$ such as the mean can be obtained from the posterior distribution corresponding to the hierarchical model proposed here. Under this model, the full conditional pdf of $\Omega$ satisfies $\pi(\Omega|Else) = \pi(\Omega|\boldsymbol{g}, G)$ and by the Diaconis-Ylvisaker theorem (Diaconis and Ylvisaker, 1979) it follows that $\Omega|Else \sim GW(\delta + 1, U + S_g), \Omega \in \mathbb{P}_G$. Because there are methods to draw samples from this distribution like the one proposed by Lenkoski (2013) and all the remaining full conditionals are standard, implementing a Gibbs sampler is straightforward. Finally, under a $\mathcal{D}\text{-W}(\boldsymbol{\alpha}, U)$ prior for $(L, D)$ and a flat prior for $\mathcal{D}$, when $n > m$, the use of the Laplace approximation permits to find an algebraic approximation to $f(\boldsymbol{y}|\mathcal{D})$; therefore, there is no need to implement any numerical integration algorithm (see Appendix). In this case, the selection of the DAG $\mathcal{D}$ is much easier from the computational point of view. To make the problem computationally tractable, the "hybrid" strategy mentioned above can be used. In this case, the EM algorithm coupled with the CSCS method (CSCS-EM) is used to estimate 15 DAG's by varying the penalty parameter of CSCS. Afterwards, once a DAG $\mathcal{D}^*$ is selected, it is used to pose a prior over $(L, D)$ after reparameterizing the hierarchical model considered here. This permits estimating the precision matrix. This method is denoted as Bayes DAG-Sel. Notice that when implementing this method; frequentist model selection is performed as an intermediate step (when estimating the 15 DAG's). In CSCS-EM and Bayes DAG-Sel, markers can be ordered according to their position in the genome.

**2.3 Simulation Study**

Two datasets were simulated in order to test the ability of the different methods proposed here to unveil the underlying conditional covariance structure of marker allele substitution effects and to estimate their concentration matrix. Notice that methods that select the sparsity pattern of the Cholesky parameter of the concentration matrix also permit estimating this matrix and consequently their ability to estimate the true conditional covariance structure can be assessed. Genotypes were simulated via a forward in time approach using the software QMSim (Sargolzaei and Schenkel, 2013). Two scenarios to simulate allele substitution effects were considered. In the first one, allele substitution effects of 300 QTL were sampled from a $N_{300}(0, \Omega^{-1})$, where $\Omega|G \sim GW(10, I_{300})$. Graph $G$ was randomly generated and then the method proposed by Lenkoski (2013), which is implemented in function rgwish of the R package BDgraph (Mohammadi and Wit, 2015), was used to draw samples of the precision matrix. The second scenario was similar, but the number of QTL (300) was smaller than sample size (380).

In the two scenarios, a single 1M chromosome was simulated, allele substitution effects were scaled to obtain an additive genetic variance of 100, and heritability was 0.5. Populations had three and four generations respectively, and the training and validation sets were defined as follows. Training sets: founders plus generations 1 and 2 (155 individuals) in scenario 1, founders and generations 1 to 3 (310 individuals) in scenario 2. Validation sets: generation 3 (45 individuals) in scenario 1 and generation 4 (70 individuals) in scenario 2. In each scenario, five replicates were simulated. The ability of the proposed models to recover the partial correlation structure was measured using the following criteria. Sensitivity, specificity, false positive rate (FPR) and the rate of correctly classified (existing or non-existing) edges (RCCE). For frequentist analyses, the penalty parameters of CONCORD-EM, GLasso-EM and CSCS-EM were chosen according to the following criteria: maximization of likelihood $L(y;\theta)$ (LIK), minimization of residual sum of squares in the validation set (RES), maximization of correlation between predicted genetic values and observed phenotypes in validation population (PA), minimization of the absolute value of the difference between true and estimated sparsity (SPA) and Bayesian information criterion (BIC). Sparsity is defined as the fraction of non-zero off-diagonal elements. In order to assess the ability to estimate the concentration matrix, the distance between true and estimated concentration matrices was measured via the Frobenious norm. Specifically, the following ratio denoted as FNR was used: $\|\hat{\Omega} - \Omega\|_F / \|\Omega\|_F$ where $\hat{\Omega}$ and $\Omega$ are the estimated and true precision matrices respectively and $\|\cdot\|_F$ denotes the Frobenious norm. Thus, the closer FNR is to zero, the smaller the distance between $\hat{\Omega}$ and $\Omega$. Finally, Pearson correlation coefficients between predicted genetic values and observed phenotypes in the validation sets (predictive abilities) and between true and predicted additive genetic values (accuracies) were computed for all methods to assess their predictive performance. Analyses were performed using existing R packages and in-house R scripts (R Core Team, 2015)

### 3 RESULTS

True sparsity was 0.1542 in scenario 1 and 0.0405 in scenario 2. In general, method GLasso-EM yielded low to moderate sensitivity and high specificity. On the other hand, sensitivities obtained from the CONCORD-EM method were markedly higher than those from Glasso-EM whereas specificities were similar. Method CSCS-EM yielded low to moderate sensitivity and its specificity values were smaller than those from the other methods. Bayes G-Sel yielded specificity and sensitivity values similar to those obtained from CONCORD-EM. Across scenarios, replicates, and methods, sensitivity was always smaller than specificity except for Glasso-EM when the tuning parameter was chosen based on criterion LIK (in this case they were equal), and Bayes DAG-Sel which yielded high sensitivities, but remarkably low specificities. Across replicates and methods, FPR varied between 0.02 and 0.13 in scenario 1 and between 0.0005 and 0.9 in scenario 2. Moreover, RCCE ranged between 0.82 and 0.91 in scenario 1 and between 0.13 and 0.98 in scenario 2. The low RCCE and high FPR values observed in scenario 2 came from the Bayes DAG-Sel method; recall that this method was implemented only in scenario 2. Bayesian information criterion always selected the sparsest precision matrix resulting in very low sensitivities and high specificities. Because of this reason, results from frequentist methods using BIC to tune the penalty parameter are not shown.

Table 1. Average sensitivity (Sens.), specificity (Spec.), false positive rate (FPR) and rate of correctly classified edges (RCCE) for the proposed methods in scenarios 1 and 2[*].

| Method[+] | Scenario 1 | | | | | Scenario 2 | | | | |
| --- | --- | --- | --- | --- | --- | --- | --- | --- | --- | --- |
| | Sens. | Spec. | FPR | RCCE | Spar. | Sens. | Spec. | FPR | RCCE | Spar. |
| Glasso-EM LIK | 0.46 (0.11) | 0.89 (0.02) | 0.11 (0.02) | 0.84 (0.01) | 0.147 (0.03) | 0.47 (0.07) | 0.95 (0.01) | 0.05 (0.01) | 0.93 (0.02) | 0.061 (0.01) |
| Glasso-EM SPA | 0.47 (0.08) | 0.89 (0.01) | 0.10 (0.02) | 0.84 (0.01) | 0.154 (0.02) | 0.34 (0.09) | 0.97 (0.01) | 0.04 (0.01) | 0.95 (0.01) | 0.042 (001) |
| Glasso-EM PA | 0.31 (0.24) | 0.89 (0.06) | 0.09 (0.03) | 0.83 (0.01) | 0.116 (0.58) | 0.41 (0.12) | 0.96 (0.01) | 0.04 (0.01) | 0.94 (0.01) | 0.053 (0.01) |
| Glasso-EM RES | 0.12 (0.02) | 0.93 (0.02) | 0.07 (0.01) | 0.83 (0.01) | 0.076 (0.02) | 0.41 (0.17) | 0.96 (0.02) | 0.04 (0.02) | 0.94 (0.01) | 0.052 (0.02) |
| CONCORD-EM LIK | 0.79 (0.02) | 0.90 (0.01) | 0.11 (0.01) | 0.88 (0.01) | 0.190 (0.01) | 0.96 (0.01) | 0.96 (0.01) | 0.04 (0.01) | 0.96 (0.01) | 0.070 (0.01) |
| CONCORD-EM SPA | 0.72 (0.02) | 0.93 (0.01) | 0.07 (0.01) | 0.90 (0.01) | 0.153 (0.01) | 0.87 (0.02) | 0.98 (0.01) | 0.02 (0.01) | 0.98 (0.001) | 0.042 (0.01) |
| CONCORD-EM PA | 0.66 (0.17) | 0.93 (0.04) | 0.07 (0.04) | 0.90 (0.01) | 0.141 (0.05) | 0.91 (0.10) | 0.97 (0.01) | 0.03 (0.02) | 0.97 (0.01) | 0.057 (0.02) |
| CONCORD-EM RES | 0.55 (0.18) | 0.95 (0.04) | 0.04 (0.03) | 0.90 (0.01) | 0.106 (0.05) | 0.85 (0.13) | 0.98 (0.02) | 0.02 (0.02) | 0.97 (0.02) | 0.049 (0.03) |
| CSCS-EM LIK | 0.41 (0.02) | 0.75 (0.01) | 0.25 (0.01) | 0.71 (0.01) | 0.269 (0.03) | 0.46 (0.08) | 0.95 (0.02) | 0.05 (0.02) | 0.94 (0.02) | 0.061 (0.01) |
| CSCS-EM SPA | 0.26 (0.02) | 0.86 (0.01) | 0.14 (0.01) | 0.79 (0.01) | 0.155 (0.02) | 0.35 (0.02) | 0.97 (0.01) | 0.03 (0.01) | 0.95 (0.01) | 0.040 (0.01) |
| CSCS-EM PA | 0.28 (0.16) | 0.83 (0.12) | 0.17 (0.12) | 0.77 (0.08) | 0.182 (0.12) | 0.46 (0.08) | 0.95 (0.02) | 0.05 (0.02) | 0.94 (0.02) | 0.061 (0.01) |
| CSCS-EM RES | 0.28 (0.13) | 0.84 (0.10) | 0.16 (0.10) | 0.78 (0.07) | 0.174 (0.11) | 0.46 (0.08) | 0.95 (0.02) | 0.05 (0.02) | 0.94 (0.02) | 0.061 (0.01) |
| Bayes G-Sel | 0.78 (0.03) | 0.90 (0.01) | 0.09 (0.01) | 0.89 (0.01) | 0.182 (0.01) | 0.92 (0.03) | 0.97 (0.01) | 0.03 (0.01) | 0.97 (0.01) | 0.056 (0.01) |
| Bayes DAG-Sel | --- | --- | --- | --- | --- | 0.95 (0.01) | 0.11 (0.02) | 0.89 (0.02) | 0.14 (0.02) | 0.887 (0.02) |

[*]Standard deviations are shown inside the parentheses.

[+]For frequentist methods, the following were the criteria used to set the tuning parameters: Maximizing the likelihood $L(y; \boldsymbol{\theta})$ (LIK), matching the true sparsity (SPA), maximizing predictive ability (PA), and minimizing the residual sum of squares in the validation set (RES).

Averages and standard deviations for the four criteria used to evaluate the ability of the proposed methods to estimate the sparsity pattern of the concentration matrix are presented in Table 1. At the individual replicate level, in scenario 1, the maximum sensitivity was 0.81 and the minimum was 0.09, while in scenario 2 the maximum was 0.96 and the minimum 0.17. Specificity ranged from 0.81 to 0.98 in scenario 1 and from 0.1 to 0.98 in scenario 2. As mentioned above, method Bayes DAG-Sel yielded very low specificities and it was implemented only in the second scenario; therefore, it is responsible for the small minimum specificity observed in this case. Removing results from Bayes DAG-Sel, the smallest specificity in scenario 2 was 0.93. As to the average (over replicates) performance, the highest sensitivities in scenario 1 were obtained from methods CONCORD-EM setting the tuning parameter according to criterion LIK and from Bayes G-Sel (Table 1). On the other hand, CONCORD-EM using the value of the tuning parameter satisfying criterion RES yielded the highest specificity and RCCE; however, CONCORD-EM using SPA and PA as criteria to tune the penalty parameter produced the same RCCE (0.90) and Bayes G-Sel yielded a very close value (0.89) (Table 1). In scenario 2, sensitivity values obtained from CONCORD-EM using criterion LIK to set the penalty parameter, Bayes G-Sel and Bayes DAG-Sel were high and similar, the highest one was obtained from CONCORD-EM. Regarding specificity, Bayes DAG-Sel had a very low average (0.11) whereas CONCORD-EM using SPA and RES to set the tuning parameter yielded the highest value and Bayes G-Sel and CSCS-EM tuning the penalty parameter using criterion SPA yielded the second highest value. As to RCCE, CONCORD-EM resulted in larger values than Glasso-EM; the highest one (0.98) was obtained under criterion SPA; notice that Bayes G-Sel and CONCORD-EM using the remaining criteria to set the penalty parameter had a very close RCCE (Table 1).

In general, frequentist approaches based on the CONCORD method (Khare et al., 2015) had a better performance than those based on the GLasso (Friedman et al. 2008) when recovering the partial correlation structure of marker effects. Method Bayes G-Sel had a good overall performance that was comparable to CONCORD-EM, especially when penalty parameter of CONCORD-EM was tuned using criterion LIK. Moreover, the overall performance of method Bayes DAG-Sel was negatively affected by the excessively high estimated sparsity that resulted in a very low specificity and a high FPR. Albeit this method exhibited a poor overall performance, it turned out to be very sensitive (Table 1). In scenario 2, the frequentist method inducing sparsity in the Cholesky parameter of the precision matrix (CSCS-EM) showed a superior overall performance as compared to Bayes DAG-Sel. Broadly speaking, methods inducing sparsity in the Cholesky parameter of the concentration matrix were outperformed by methods inducing sparsity in the concentration matrix directly; however, it is worth mentioning that in scenario 2, results from CSCS-EM were very similar to those from Glasso-EM (Table 1). Method CSCS-EM featured low sensitivity, but its performance in the remaining parameters varied across scenarios; therefore, it is worth discussing them separately.  In scenario 1, specificity was reasonably high, FPR values were larger than those obtained with Glasso-EM, CONCORD-EM and Bayes G-Sel, and RCCE values were smaller than those obtained from the other methods. In contrast, in scenario 2, specificity was high, FPR values were low and comparable to those yielded by the other methods, and RCCE was high. In this scenario, criteria LIK, SPA and RES selected the same DAG in all replicates.

Table 2. Average (over replicates) predictive abilities (APA), accuracies in training (AAT) and validation sets (AAV), and ratio of Frobenious norms (AFNR)*.

| Method[+] | Scenario 1 | | | | Scenario 2 | | | |
|---|---|---|---|---|---|---|---|---|
| | APA | AAV | AAT | AFNR | APA | AAV | AAT | AFNR |
| GLasso-EM LIK | 0.63 | 0.83 | 0.86 | 0.65 | 0.63 | 0.80 | 0.80 | 0.57 |
| | (0.10) | (0.06) | (0.04) | (0.06) | (0.01) | (0.06) | (0.06) | (0.01) |
| GLasso-EM SPA | 0.64 | 0.84 | 0.86 | 0.65 | 0.61 | 0.75 | 0.81 | 0.64 |
| | (0.10) | (0.07) | (0.04) | (0.04) | (0.02) | (0.04) | (0.04) | (0.01) |
| GLasso-EM PA | 0.65 | 0.84 | 0.86 | 0.81 | 0.63 | 0.77 | 0.81 | 0.59 |
| | (0.11) | (0.07) | (0.05) | (0.04) | (0.01) | (0.05) | (0.05) | (0.02) |
| GLasso-EM RES | 0.64 | 0.85 | 0.87 | 0.84 | 0.63 | 0.77 | 0.81 | 0.60 |
| | (0.11) | (0.07) | (0.04) | (0.004) | (0.01) | (0.05) | (0.05) | (0.04) |
| CONCORD-EM LIK | 0.58 | 0.83 | 0.87 | 0.13 | 0.64 | 0.81 | 0.81 | 0.06 |
| | (0.10) | (0.06) | (0.03) | (0.01) | (0.06) | (0.04) | (0.05) | (0.002) |
| CONCORD-EM SPA | 0.58 | 0.84 | 0.88 | 0.17 | 0.64 | 0.80 | 0.80 | 0.09 |
| | (0.10) | (0.07) | (0.03) | (0.01) | (0.06) | (0.04) | (0.04) | (0.005) |
| CONCORD-EM PA | 0.59 | 0.83 | 0.88 | 0.21 | 0.64 | 0.81 | 0.81 | 0.07 |
| | (0.09) | (0.06) | (0.03) | (0.11) | (0.06) | (0.04) | (0.05) | (0.03) |
| CONCORD-EM RES | 0.58 | 0.83 | 0.87 | 0.27 | 0.64 | 0.80 | 0.81 | 0.09 |
| | (0.09) | (0.08) | (0.03) | (0.11) | (0.06) | (0.05) | (0.05) | (0.003) |
| CSCS-EM LIK | 0.52 | 0.74 | 0.81 | 0.66 | 0.63 | 0.79 | 0.80 | 0.48 |
| | (0.10) | (0.15) | (0.07) | (0.02) | (0.04) | (0.04) | (0.05) | (0.01) |
| CSCS-EM SPA | 0.51 | 0.74 | 0.81 | 0.70 | 0.62 | 0.77 | 0.80 | 0.51 |
| | (0.10) | (0.14) | (0.07) | (0.01) | (0.03) | (0.05) | (0.05) | (0.03) |
| CSCS-EM PA | 0.52 | 0.74 | 0.81 | 0.69 | 0.63 | 0.79 | 0.80 | 0.48 |
| | (0.10) | (0.15) | (0.07) | (0.04) | (0.04) | (0.04) | (0.05) | (0.01) |
| CSCS-EM RES | 0.52 | 0.74 | 0.81 | 0.74 | 0.63 | 0.79 | 0.80 | 0.48 |
| | (0.10) | (0.15) | (0.07) | (0.14) | (0.04) | (0.04) | (0.05) | (0.01) |
| Bayes G-Sel | 0.55 | 0.77 | 0.83 | 0.73 | 0.63 | 0.82 | 0.84 | 0.73 |
| | (0.05) | (0.02) | (0.05) | (0.04) | (0.06) | (0.07) | (0.04) | (0.05) |
| Bayes DAG-Sel | ---- | ---- | ---- | ---- | 0.53 | 0.70 | 0.74 | 0.74 |
| | | | | | (0.03) | (0.06) | (0.01) | (0.20) |

*Standard deviations are shown inside the parentheses.
[+]For frequentist methods, the following were the criteria used to set the tuning parameters: Maximizing the likelihood $L(y; \boldsymbol{\theta})$ (LIK), matching the true sparsity (SPA), maximizing predictive ability (PA), and minimizing the residual sum of squares in the validation set (RES).

Regarding the estimated precision matrices, CONCORD-EM exhibited an outstanding accuracy because average FNR (AFNR) values were much lower than those obtained from the other methods (Table 2). In scenario 1, CONCORD-EM with the penalty parameter tuned according to criteria LIK had the smallest (best) AFNR value whereas the highest was obtained from CSCS-EM using RES to tune the penalty parameter. In scenario 2, the smallest AFNR was obtained from method CONCORD-EM using LIK to set the penalty parameter and the highest from Bayes DAG-Sel (Table 2).

Predictive abilities and accuracies of predicted additive genetic values did not show differences as marked as those found for the ability to estimate $\Omega$ (Table 2). This result suggests that methods estimating the precision matrix appropriately do not necessarily exhibit a markedly superior predictive performance.

In that regard, notice that in scenario 1, the highest predictive abilities and accuracies were attained by Glasso-EM (Table 2) which did not have the best performance estimating $\Omega$ (Table 1, Table 2). In scenario 2, the situation was somewhat different. There, CONCORD-EM, the method that better estimated the partial correlation structure and the concentration matrix, had a slightly better APA than the other models. Moreover, Bayes DAG-Sel which did the poorest job in estimating the sparsity pattern of $\Omega$, also had the worst predictive performance, but predictive abilities and accuracies obtained from this method were not extremely low. Bayes G-Sel yielded the highest AAT and AAV values (Table 2).

## 4 DISCUSSION
### 4.1 Comments on the Methods

In this study, the problem of estimating the partial correlation structure and the precision matrix of allele substitution effects in genome-wide prediction was addressed. To this end, the theory of graphical models to perform inverse covariance estimation when the sparsity pattern of the precision matrix is unknown was adapted to a linear regression model in which phenotypes are regressed on mappings of marker genotypes. The methods proposed here cope with the problem of estimating the precision matrix of an unobservable $m$-dimensional random vector using a single $n$-dimensional observable vector containing phenotypic information and a design matrix containing genotypic information via GCGM and Gaussian DAG models. Specifically, three frequentist methods, Graphical Lasso (Friedman et al., 2008), CONCORD (Khare et al., 2015) and CSCS (Khare et al., 2016) were adapted to perform model selection/DAG selection in genome-wide prediction using the EM algorithm (Dempster et al., 1977) giving rise to GLasso-EM, CONCORD-EM and CSCS-EM methods. On the Bayesian side, the flexibility of hierarchical Bayesian modeling combined with MCI permitted to tackle the problem. In addition, when $m > n$, the Laplace approximation facilitated DAG selection because it avoided the use of numerical integration methods. Once a graph is selected, under the hierarchical models considered here, the full conditional distribution of the precision matrix or its Cholesky parameters pertain to the same family of the prior and as a consequence, implementing a Gibbs sampler is straightforward because direct sampling from these distributions is feasible and all the remaining full conditionals are standard.

It has to be considered that when $n > m$, for Bayes G-Sel the use of the Laplace approximation permitted a faster partial correlation graph selection under the particular algorithms and software used in this study. Specifically, the Monte Carlo method of Atay-Kayis and Massam (2005) was used to compute the normalizing constant of the G-Wishart distribution and the method of Lenkoski (2013) was used to draw samples from this distribution. These two methods are implemented in the R package BDgraph (Mohammadi and Wit, 2015). For example, for 300 variables, approximation of the normalizing constant of a G-Wishart distribution using the method of Atay-Kayis and Massam (2005) and 5000 samples took around 22 seconds. On the other hand, drawing a single sample from the same distribution using the method of Lenkoski (2013) took around 3.5 seconds; this is why this approach was much slower. However, this result may not hold when using different algorithms and/or software. In addition, the implementation of the Monte Carlo method of Atay-Kayis and Massam (2005) in the aforementioned R package exhibited some issues. It yielded the same value of the log-normalizing constant for substantially different graphs. We believe that it is due to numerical problems. Because of this reason, results presented in Table 1 for scenario 2 were obtained using a full MCI approach.

The methods developed here permit learning the partial correlation network of the effects of molecular markers considered in the model on a given phenotype, estimating their precision matrix, and accounting for correlated marker effects in genome-wide prediction models. On the Bayesian side, this last goal is automatically achieved when estimating the precision matrix. On the frequentist side, it can be easily achieved by plugging $\widehat{\Omega}$ and $\hat{\sigma}^2$ in the mixed model equations corresponding to the linear model described in Equation 3 and solving them to obtain the empirical BLUP of $\boldsymbol{g}$: $\widehat{\boldsymbol{g}} = \left(W'W + \hat{\sigma}^2 \widehat{\Omega}\right)^{-1} W'\boldsymbol{y}$ (Henderson, 1950; 1963).

It has been reported that highly connected nodes are relevant in different biological networks like gene co-expression (Carter et al., 2004) and protein-protein interaction networks (Han et al., 2004). For example, using gene expression data, the so called "hub genes" (those that are highly connected in a given graph) have been found to play important roles in breast cancer development (Peng et al., 2009; Khare et al., 2015). Thus, our methods are useful to simultaneously studying the "genetic architecture" of a given trait and predicting additive genetic values taking into account the inferred conditional covariance structure which may lead to more accurate predictions. Therefore, identification of "hub" SNP's using the proposed methods may be helpful in identifying regions of the genome affecting a given trait; for example, these analyses may be used to complement results from genome-wide association studies or gene set analyses.

So far, in the realm of quantitative genetics, there have been few published papers addressing the issue of accounting for correlated marker effects. Gianola et al. (2003) and Yang and Tempelman (2012) considered the estimation of a non-diagonal covariance matrix of marker effects, that is, they considered marginally correlated effects and imposed restrictive marginal covariance structures dictated by the physical location of markers in such a way that only nearby markers were correlated. Gianola et al. (2003) proposed a series of frequentist and Bayesian methods that assumed only intra-chromosome correlations and some of their methods required equidistant markers. On the other hand, Yang and Tempelman (2012) proposed a Bayesian first-order nonstationary antedependence model. The idea of spatial correlation makes sense because nearby markers could be linked to the same gene or set of genes having a true effect on the trait. However, due to the complex interactions between genes and gene products, there could be correlations between effects of markers located far away in the same chromosome or even in different chromosomes. In Martínez et al. (2016a; 2016b), models for sparse estimation of the covariance or precision matrix using domain-specific knowledge to build the undirected graph *G* encoding the covariance (marginal or partial) structure were developed. These methods are flexible because they allow considering correlations or partial correlations that are not explained by physical position. The methods developed here are even more flexible because they allow estimation of the precision matrix without assuming a known pattern of zeros.

    Scalability is an issue for most of the existing methods used to estimate the sparsity pattern of either the precision matrix or its Cholesky parameter. So far, the number of variables (nodes) considered in simulation and real data analyses vary from tens or a few hundreds (Friedman et al., 2008; Bickel and Levina, 2008; Wang, 2012) to a few thousands for the most efficient methods like SPACE and CONCORD (Peng et al., 2009; Khare et al., 2015). Thus, there is a need to develop strategies that permit to implement these methods in larger datasets.

    When inducing sparsity in the precision matrix directly, for the special case of decomposable graphs, there is no need for MCI due to the fact that the normalizing constant of the G-Wishart distribution has closed form (Letac and Massam, 2007, Rajaratnam et al., 2008) and consequently, for full rank models, after plugging the Laplace approximation to the inner integral in Equation 5 as explained above, the external integral can be found in closed form. On the other hand, sampling is faster for this sort of graphs and consequently, evaluating this external integral is faster for the non-full rank case. Hence, a compromise between generality and computational efficiency would be to restrict the problem by considering the space of certain families of decomposable graphs. As discussed above, the assumption of spatial correlation has been used in genome-wide prediction and it has biological justification (Gianola et al., 2003; Yang and Templeman, 2012). Assuming that only markers within a given distance are partially correlated induces a banded or a differentially banded concentration matrix, and these structures correspond to decomposable graphs. In addition, the nested Lasso penalty (Levina et al., 2008) induces a banded structure on the Cholesky parameter of the precision matrix and selects the bandwidth for each row. In genome-wide prediction, the order dictated by the physical location of markers can be regarded as a natural ordering. Thus, this method could be adapted to genome-wide prediction by following a similar approach to the one used here. Similarly, other Bayesian approaches to perform model selection can be easily adapted thanks to the flexibility of hierarchical modelling. For example, Wang (2012) proposed the Bayesian graphical Lasso and used a thresholding approach to select the sparsity pattern of the precision matrix. This model

poses independent Laplace priors to the off-diagonal entries of Ω and independent exponential priors to its diagonal entries (Wang, 2012).

For the CONCORD method, Khare et al. (2015) proved model selection consistency in high dimensional settings (i.e., when $n$ and $m$ increse to infinity) under four regularity conditions. For the thresholding approach, Bickel and Levina proved consistency in the operator norm under Gaussianity or sub-Gaussianity if $(\log p)/n \rightarrow 0$. These proofs were done for the standard scenario where the sparsity pattern of the concentration matrix of observable random variables is estimated. Therefore, studying large sample properties of the estimators obtained with the methods developed here remains an open problem.

Finally, it is worth mentioning that extension of these methods to the multiallelic case is straightforward. The only change needed is to modify the design matrix in order to include the effects of several alleles per locus. One way to do it is to proceed as shown in Martínez et al. (2016a; 2016b). Then, after redefining the design matrix, no further changes are required to implement the proposed methods in the multiallelic case. However, notice that when dealing with multiallelic loci, the dimensionality increases.

## 4.2 Simulation Results

As stated above, to our knowledge, this is the first study addressing graphical model selection/DAG selection in the context of genome-wide prediction; therefore, there are no comparable studies to be discussed. Some studies using model selection methods to estimate conditional covariance in the standard setting (i.e., estimating the precision matrix of observable random vectors using a sample of size $N$) have reported satisfactory performance in estimating the true partial correlation structure using simulated data. The following are some examples, Bickel and Levina (2008): thresholding , Peng et al. (2009): SPACE, Wang (2012): Bayesian graphical Lasso, Khare et al. (2015): CONCORD. Not all the authors used the same metrics to assess how well the partial correlation structure was estimated. Some of the metrics used in these studies were: number of correctly detected nodes, area under the curve of ROC curves, norms of the difference between estimated and true concentration matrices, sensitivity, specificity and Matthews correlation coefficient.

The overall performance of most of the proposed methods was satisfactory, because RCCE, sensitivity and specificity were reasonably high and FPR was low. The rate of correctly classified edges is a criterion that combines the ability to identify true positives and true negatives; therefore, the larger the sensitivity and specificity, the larger the RCCE. In fact, it can be easily shown that RCCE is equal to one if and only if sensitivity and specificity equal one as well. Thus, RCCE can be seen as a good summary of the overall graph selection performance. However, it has to be considered that due to the low sparsity values considered in this simulation study, methods with a low sensitivity can have a relatively high RCCE; consequently, in this case sensitivity values also have to be considered. That being said, notice that except for Bayes DAG-Sel, all methods yielded a RCCE larger or equal than 0.71 (Table 1). Furthermore, methods inducing sparsity in the precision matrix directly tended to perform better than those inducing sparsity in its Cholesky parameter.

According to the results from these simulations, it seems that for frequentist methods, the criterion used to tune the penalty parameter affects their ability to estimate the pattern of zeros of the precision matrix, especially sensitivity. The only exception was CSCS-EM in scenario 2 where three of the four criteria selected the same DAG and yielded the same sparsity pattern in $\Omega$ in all replicates. In general, criterion SPA always had a good performance; in particular, within each method, this criterion always produced the highest RCCE. This suggests that when the tuning parameter is chosen on the basis of a desired sparsity $\varphi$, this approach might perform well as far as $\varphi$ is close to the true sparsity. As to the methods inducing sparsity directly in $\Omega$, criterion LIK had an acceptable performance in the two scenarios, for methods CONCORD-EM and CSCS-EM it always had the highest sensitivity as well as for GLasso-EM in scenario 2, while for GLasso-EM in scenario 1, its sensitivity was the second largest and it was very close to the largest one (Table 1). Moreover, tuning the penalty parameter using criterion PA did not result in noticeable differences in predictive ability or accuracy of predicted genetic values (Table 2). In fact, in many cases, APA values differed just in the third or fourth decimal positions. Thus, from the predictive point of view, this method to tune the penalty parameters did not show superiority. Notwithstanding, this criterion had a good performance when estimating the sparsity pattern of $\Omega$; its only drawback was its moderate sensitivity in scenario 1. When inducing sparsity in the Cholesky parameter of $\Omega$, the performance of these criteria was slightly different. Maximizing the likelihood $L(\boldsymbol{y};\boldsymbol{\theta})$ was still the approach yielding the highest sensitivity; however, the drop in specificity and RCCE compared with the other approaches was more evident and FPR was much higher in scenario 1 (Table 1). In addition, although criteria PA and RES had smaller sensitivities than SPA, the difference was not as big as when inducing sparsity in $\Omega$, and these two criteria performed better in terms of specificity, FPR and RCCE. Finally, AFNR, a criterion informing about "similarity" between estimated and true concentration matrices, showed a marked superiority of CONCORD-EM coupled with the method of Hastie et al. (2009) (Table 2). Because AFNR expresses the Frobenious norm of the difference between estimated and true concentration matrices relative to the Frobenious norm of the true concentration matrix, it is comparable across scenarios. For all methods, AFNR values from scenario 2, where the number of phenotypes was larger than the number of marker loci and true sparsity was smaller, were smaller than those from scenario 1. This indicates that when the ratio of phenotypic observations to covariance parameters increased, the estimates of the precision matrix were closer to the true one. With the exception of CONCORD-EM combined with the method of Hastie et al. (2009) in scenario 2, methods yielding estimated precision matrices closer to the true one did not necessarily have the best predictive performance (Table 2). Moreover, excluding CONCORD-EM, methods recovering the sparsity pattern of the precision matrix more accurately did not exhibit the same superiority in terms of the "similarity" between estimated and true precision matrices as measured by the Frobenious norm. The most conspicuous example was Bayes G-Sel (Table 1, Table 2). Therefore, although the pattern of zeros inferred from some methods was closer to the true one, the estimated non-zero entries of the precision matrix were more distant from the true values resulting in a large AFNR. Consequently, if method A infers the partial correlation network more accurately than method B, it does not necessarily yield a more accurate estimate of the precision matrix. However, it is worth mentioning again, that results from this study showed an outstanding performance of methods based on CONCORD when estimating the precision matrix. These results also indicate that predictive performance is not necessarily improved when the partial correlation structure is inferred with better accuracy and that methods showing marked differences in their ability to estimate the precision matrix accurately may exhibit a very similar predictive performance.

## 5 FINAL REMARKS

Unlike methods developed in Martínez et al. (2016a), the statistical methods developed in this study do not require previous definition of the sparsity pattern of the concentration matrix because they have the ability to learn it from observed data (phenotypes and genotypes). These methods fulfill two objectives, to estimate the partial correlation structure and the concentration matrix of marker effects, and to incorporate it in the prediction of additive genetic values. Therefore, they can be used as predictive machines and as tools to learn about the covariation of effects of pairs of loci on a given phenotype conditioned on the effects of all the other loci considered in the model. Results from the simulation study performed here suggest that some of these methods can recover the partial correlation structure and estimate the concentration matrix with satisfactory accuracy. Therefore, they are useful tools to learn about the underlying biology of a given trait because they help to understand relationships between different regions of the genome in terms of the partial correlations of their effects on that trait. Furthermore, these methods permit the incorporation of that information in the prediction of additive genetic values. Although this study focused on biallelic loci, the fact that the proposed methods are easily extended to the multiallelic case permits to implement them in studies considering multiallelic loci.

## ACKNOWLEDGEMENTS

C.A. Martínez thanks Fulbright Colombia and "Departamento Adiministrativo de Ciencia, Tecnología e Innovación" COLCIENCIAS for supporting his PhD and Master programs at the University of Florida through a scholarship, and Bibiana Coy for her love, support, and constant encouragement.

# APPENDIX
# Details on the Use of the Laplace Approximation to Find Marginal Likelihoods

**Full Rank Model Case**

The Laplace approximation permits to find integrals of the form $I = \int_{\mathbb{R}^p} q(\boldsymbol{\theta})e^{nh(\boldsymbol{\theta})}d\boldsymbol{\theta}$, where $q$ and $h$ are smooth functions of $\boldsymbol{\theta}$ and $h$ has a unique maximum at $\widehat{\boldsymbol{\theta}}$. It has the form (Ghosh et al., 2006): $I = \exp\left(nh(\widehat{\boldsymbol{\theta}})\right)(2\pi)^{p/2}n^{-p/2}|\Delta_h(\widehat{\boldsymbol{\theta}})|^{-1/2}q(\widehat{\boldsymbol{\theta}})(1 + O(n^{-1}))$, where $p = \dim(\boldsymbol{\theta})$ and $|\Delta_h(\widehat{\boldsymbol{\theta}})|$ is the determinant of the Hessian matrix of $-h$ evaluated at $\widehat{\boldsymbol{\theta}}$. If $nh(\boldsymbol{\theta})$ is taken to be the log-likelihood, then $\widehat{\boldsymbol{\theta}}$ is the MLE of $\boldsymbol{\theta}$. It implies that the linear model presented in Equation 3 has to be of full rank, and this is why this approximation can be applied only when $n \geq m$, because this condition is necessary for matrix $W$ to be of full column rank. The following are the details of the derivation of the Laplace approximation to

$$I_1 := \int_{\mathbb{R}^{ms}} \int_{\mathbb{R}_+} f(\boldsymbol{y}|\boldsymbol{g}, \sigma^2, W)\pi(\boldsymbol{g}|\Omega)\pi(\sigma^2)\, d\sigma^2 d\boldsymbol{g}.$$

The first step is to rearrange the integrand as follows:

$$I_1 = \int_{\mathbb{R}^{ms+1}} q(\boldsymbol{\theta}^*)\exp(nh(\boldsymbol{\theta}^*))\, d\boldsymbol{\theta}^*,$$

where $\boldsymbol{\theta}^* := (\boldsymbol{g}, \sigma^2)$, $nh(\boldsymbol{\theta}^*) := \ln(f(\boldsymbol{y}|\boldsymbol{g}, \sigma^2, W))$ and $q(\boldsymbol{\theta}^*) := \pi(\boldsymbol{g}|\Omega)\pi(\sigma^2)$. The sampling distribution is a $MVN(W\boldsymbol{g}, \sigma^2 I)$; therefore, following standard results from linear models theory, if $W$ is of full column rank then, $\widehat{\boldsymbol{g}} = (W'W)^{-1}W'\boldsymbol{y}$ is the MLE of $\boldsymbol{g}$, and $\widehat{\sigma}^2 = \frac{\|\boldsymbol{y}-W\widehat{\boldsymbol{g}}\|^2}{n} = \frac{\boldsymbol{y}'(I-H_W)\boldsymbol{y}}{n} = \frac{(n-r)}{n}S^2$ is the MLE of $\sigma^2$, where $S^2 = \frac{\boldsymbol{y}'(I-H_W)\boldsymbol{y}}{n-r}$ is the least squares estimator of $\sigma^2$, $r = rank(W'W) = m$ and $H_W = W(W'W)^{-1}W$ is the projection matrix onto the column space of $W$.

For the model considered here, the Hessian matrix of the log-likelihood is:

$$H = \frac{1}{\sigma^2}\begin{pmatrix} -W'W & \frac{1}{\sigma^2}(W'W\boldsymbol{g} - W'\boldsymbol{y}) \\ Sym & \frac{1}{\sigma^2}\left(\frac{n}{2} - \frac{(\boldsymbol{y}-W\boldsymbol{g})'(\boldsymbol{y}-W\boldsymbol{g})}{\sigma^2}\right) \end{pmatrix},$$

thus, matrix $\Delta_h(\widehat{\boldsymbol{\theta}}^*)$ is:

$$\begin{pmatrix} \frac{W'W}{(n-r)S^2} & \left(\frac{n}{(n-r)S^2}\right)^2 (W'W(W'W)^{-1}W'\boldsymbol{y} - W'\boldsymbol{y}) \\ sym & \frac{n^2}{2\left((n-r)S^2\right)^2} \end{pmatrix}$$

$$= \begin{pmatrix} \frac{W'W}{SSR} & 0 \\ 0 & \frac{n^2}{2SSR^2} \end{pmatrix},$$

therefore;

$$|\Delta_h(\widehat{\boldsymbol{\theta}}^*)|^{1/2} = \frac{n}{SSR^{(m+2)/2}} \left(\frac{|W'W|}{2}\right)^{\frac{1}{2}} := \Delta$$

$$q(\widehat{\boldsymbol{\theta}}^*) = \pi(\widehat{\boldsymbol{g}}|\Omega)\pi(\widehat{\sigma}^2)$$

$$= (2\pi)^{-m/2}|\Omega|^{1/2}\exp\left(-\frac{1}{2}(\widehat{\boldsymbol{g}}'\Omega\widehat{\boldsymbol{g}})\right)\frac{(\tau^2)^{v/2}}{\Gamma\left(\frac{v}{2}\right)2^{v/2}}(\widehat{\sigma}^2)^{-(v/2+1)}\exp\left(\frac{-\tau^2}{2\widehat{\sigma}^2}\right),$$

thus

$$\int_{\mathbb{R}^{mS}}\int_{\mathbb{R}_+} \pi(\boldsymbol{g}|\Omega)\pi(\sigma^2)e^{\ln f(\boldsymbol{y}|\boldsymbol{g},\sigma^2,W)}\,d\sigma^2 d\boldsymbol{g} \approx \frac{1}{\Delta}(2\pi)^{-(n+1)/2}\left(\frac{SSR}{n}\right)^{-(n+v+2)/2}\exp\left(-\frac{n}{2}\right)$$

$$\times \frac{(\tau^2)^{v/2}}{\Gamma\left(\frac{v}{2}\right)2^{v/2}}\exp\left(-\frac{n\tau^2}{2SSR}\right)n^{-(m+1)/2}|\Omega|^{1/2}\exp\left(-\frac{1}{2}(\widehat{\boldsymbol{g}}'\Omega\widehat{\boldsymbol{g}})\right)$$

$$= \frac{1}{\Delta}(2\pi)^{-(n+1)/2}\left(\frac{SSR}{n}\right)^{-(n+v+2)/2}\exp\left(-\frac{n}{2}\right)\frac{(\tau^2)^{v/2}}{\Gamma\left(\frac{v}{2}\right)2^{v/2}}\exp\left(-\frac{n\tau^2}{2SSR}\right)n^{-(m+1)/2}$$

$$\times |\Omega|^{1/2}\exp\left(-\frac{1}{2}(\widehat{\boldsymbol{g}}'\Omega\widehat{\boldsymbol{g}})\right).$$

$$:= \frac{K}{\Delta}|\Omega|^{1/2}\exp\left(-\frac{1}{2}(\widehat{\boldsymbol{g}}'\Omega\widehat{\boldsymbol{g}})\right),$$

where

$$K := (2\pi)^{-(n+1)/2}\left(\frac{SSR}{n}\right)^{-(n+v+2)/2}\exp\left(-\frac{n}{2}\right)\frac{(\tau^2)^{v/2}}{\Gamma\left(\frac{v}{2}\right)2^{v/2}}\exp\left(-\frac{n\tau^2}{2SSR}\right)n^{-(mS+1)/2}$$

$$SSR := \boldsymbol{y}'(I - H_W)\boldsymbol{y} = (n-r)S^2$$

$$\widehat{\boldsymbol{g}} := (W'W)^{-1}W'\boldsymbol{y}.$$

Notice that $K$ and $\Delta$ are the same for all the models considered here; therefore, these expressions do not affect model selection. Now, using the Laplace approximation to $I_1$ it follows that:

$$f(\boldsymbol{y}|G) \approx \frac{K}{\Delta}\int_{\mathbb{P}_G} \pi(\Omega)|\Omega|^{1/2}\exp\left(\frac{-1}{2}\widehat{\boldsymbol{g}}'\Omega\widehat{\boldsymbol{g}}\right)d\Omega$$

$$\propto \int_{\mathbb{P}_G} |\Omega|^{(1+\delta)/2}\exp\left(\frac{-1}{2}tr\big(\Omega(\widehat{\boldsymbol{g}}\widehat{\boldsymbol{g}}' + U)\big)\right)d\Omega$$

which is nothing but the normalizing constant of a $GW(1+\delta, \widehat{\boldsymbol{g}}\widehat{\boldsymbol{g}}' + U)$ distribution.

For Bayes DAG-Sel, after reparameterizing in terms of the modified Cholesky decomposition of $\Omega$, defined as $\Omega = LD_*^{-1}L'$

$$f(\mathbf{y}|\mathcal{D}) \approx \frac{K}{\Delta} \int_\mathbb{D} \int_{L_\mathcal{D}} \pi(L,D) \left(\prod_{j=1}^{m} |D_{*ii}|^{1/2}\right) \exp\left(\frac{-1}{2}\hat{\mathbf{g}}'LD_*^{-1}L'\hat{\mathbf{g}}\right) dLdD$$

$$\propto \int_\mathbb{D} \int_{L_\mathcal{D}} \left(\prod_{j=1}^{m} |D_{*ii}|^{-(1+\alpha_j)/2}\right) \exp\left(-\frac{1}{2}tr\left((LD_*^{-1}L')(\hat{\mathbf{g}}\hat{\mathbf{g}}' + U)\right)\right) dLdD$$

$$= \prod_{j=1}^{p} \frac{\Gamma\left(\frac{\tilde{\alpha}_j - |pa_j|}{2} - 1\right) 2^{\tilde{\alpha}_j/2 - 1} \pi^{\frac{|pa_j|}{2}} |\tilde{U}^{<j}|^{\frac{\tilde{\alpha}_j - |pa_j| - 3}{2}}}{|\tilde{U}^{\leq j}|^{\frac{\tilde{\alpha}_j - |pa_j|}{2} - 1}}$$

where $\tilde{U} = \hat{\mathbf{g}}\hat{\mathbf{g}}' + U$ and $\tilde{\alpha}_j = \alpha_j + 1 \ \forall \ j = 1,2,\ldots,m$ and $\mathbb{D}$ is the space of diagonal matrices of dimension $m \times m$ with positive diagonal elements. The last equality follows by applying Equation 2 after noticing that the integrand is the kernel of a DAG-W$(\tilde{\alpha}, \tilde{U})$ distribution and consequently the integral is equal to the corresponding normalizing constant.

**Non-Full Rank Model Case**

For the case $m > n$, the Laplace approximation can be applied by modifying the prior and redefining $nh(\boldsymbol{\theta})$. If $\mathbf{g}|\Omega, \sigma^2 \sim MVN(\mathbf{0}, \sigma^2\Omega^{-1})$ and $nh(\boldsymbol{\theta}) \coloneqq \ln[f(\mathbf{y}|\mathbf{g}, W, \sigma^2)\pi(\mathbf{g}|\Omega, \sigma^2)]$ then finding the marginal likelihood presented in Equation 5 amounts to computing the expectation of a function of $\Omega$ with respect to a $GW(\delta + 1, U)$ distribution. This expectation can be computed via vanilla MCI. This modification is made in order to "beat the singularity" of $W'W$. With this setup, the problem is finding:

$$f(\mathbf{y}|G) = \int_{\mathbb{P}_G} \pi(\Omega|G) \left(\int_{\mathbb{R}^m} \int_{\mathbb{R}_+} f(\mathbf{y}|\mathbf{g}, \sigma^2, W)\pi(\mathbf{g}|\Omega, \sigma^2)\pi(\sigma^2)\, d\sigma^2 d\mathbf{g}\right) d\Omega$$

In order to find the Laplace approximation to the inner integral

$$I_2 \coloneqq \int_{\mathbb{R}^m} \int_{\mathbb{R}_+} f(\mathbf{y}|\mathbf{g}, \sigma^2, W)\pi(\mathbf{g}|\Omega, \sigma^2)\pi(\sigma^2)\, d\sigma^2 d\mathbf{g}$$

Basically, the problem of $W'W$ being singular is that it is not possible to find the Laplace approximation by defining $nh(\boldsymbol{\theta})$ as the log-sampling distribution because the inverse of this matrix does not exist and this inverse is required. Thus, $nh(\boldsymbol{\theta})$ is now defined as $\ln[f(\mathbf{y}|\mathbf{g}, W, \sigma^2)\pi(\mathbf{g}|\Omega, \sigma^2)]$, whit this definition, after factoring expressions involving $\sigma^2$, the matrix that has to be inverted is $A \coloneqq W'W + \Omega$. By standard results from matrix algebra, this matrix is invertible disregarding the rank of $W$ because $\Omega$ is positive definite and $W'W$ is symmetric. Following analogous steps to those shown for the full rank model, it can be shown that the Laplace approximation to $I_2$ is proportional to

$$|\Omega|^{1/2}|A|^{-1/2}\exp(-n\tau^2/(2S_*^2))\,(S_*^2)^{-(v+n)/2}$$

where $S_*^2 := \boldsymbol{y}'(I - WA^{-1}W')\boldsymbol{y}$; thus,

$$f(\boldsymbol{y}|G) \approx K^* \int_{\mathbb{P}_G} \pi(\Omega|G)\,|\Omega|^{1/2}|A|^{-1/2}\exp(-n\tau^2/(2S_*^2))\,(S_*^2)^{-(v+n)/2}d\Omega$$

$$\propto \int_{\mathbb{P}_G} |A|^{-1/2}\exp(-n\tau^2/(2S_*^2))\,(S_*^2)^{-(v+n)/2}\,|\Omega|^{(\delta+1)/2}\exp(-tr(\Omega U)/2)\,d\Omega$$

$$\propto E\!\left[|A|^{-1/2}\exp(-n\tau^2/(2S_*^2))\,(S_*^2)^{-(v+n)/2}\right]$$

where $K^*$ is an expression not depending on $\Omega$ and the expectation is taken with respect to a $GW(\delta+1,U)$ distribution. Finally, notice that when the prior for $\boldsymbol{g}$ is a $MVN(\boldsymbol{0},\sigma^2\Omega^{-1})$, then $\Omega|Else \sim GW(\delta+1, U + S_g/\sigma^2)$. Thus, as in the full rank model case, implementing a Gibbs sampler is straightforward.